# Design of a Virtual Component Neutral Network-on-Chip Transaction Layer

Philippe Martin
Arteris S.A.
6 Parc Ariane – 78284 Guyancourt - France
philippe.martin@arteris.com

## Abstract

*Research studies have demonstrated the feasibility and advantages of Network-on-Chip (NoC) over traditional bus-based architectures but have not focused on compatibility communication standards. This paper describes a number of issues faced when designing a VC-neutral NoC, i.e. compatible with standards such as AHB 2.0, AXI, VCI, OCP, and various other proprietary protocols, and how a layered approach to communication helps solve these issues.*

## 1. NoC Layered Architecture

A true Network-On-Chip uses a layered communication approach, similar to the OSI model which is at the foundation of LANs and WAN, but simpler. Arteris NoC defines transaction, transport, and physical layers.

The NoC transaction layer defines communication primitives available to IP blocks that are plugged into the NoC. A Network Interface Unit (NIU) is responsible for converting the foreign IP protocol to the NoC transaction layer.

The transport layer defines information format and transport rules between NIUs. Arteris NoC packet-based transport layer defines how packets are routed, quality of service used for prioritizing packets, etc. The transport layer is completely transaction unaware, and conversely, transaction level is transport unaware (for example, wormhole or store-and-forward packet handling makes no difference at the transaction level).

The physical layer defines how packets are physically transmitted – much like the Ethernet defines the MII, 10Mb/s, 1Gb/s physical interfaces. Again, the physical layer is independent from transaction and transport layers.

Layer independence enables a separate optimization of NoC features: the transaction layer focuses on compatibility with existing IP blocks and busses, transport layer focuses on quality of service and scalability, physical layers on implementation related features such as achieving raw bandwidth, matching clocks, off-chip communication, etc.

## 2. VC compatibility handling

To create System-On-Chip (SoC), VCs with mixed sockets ideally plug into an interconnect according to Fig 1:

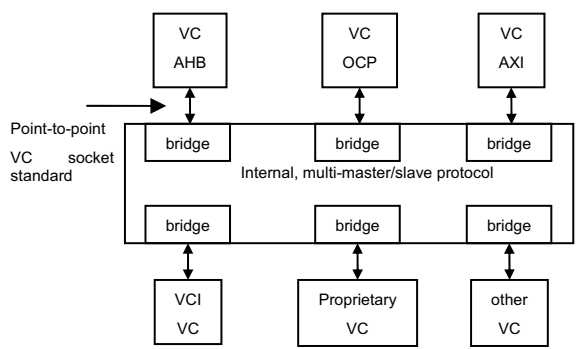

**Fig 1: Ideal system with mixed VC protocols.**

Interconnects use a multi-master and multi-slave protocol and include bridges to the various VC socket protocols.

In practice, the intertwining of transaction, transport and physical levels within standard interconnects, and the interconnect flexibility necessary to handle many application designs makes this ideal approach very difficult. Designs mixing different VC standards look more like Fig 2, where the interconnect has its own reference socket standard. Bridges to the reference standard are used plug the IP blocks (an alternative is to redesign the IP to support directly the interconnect reference standard, but this is time-consuming and prone to errors, and not possible for off-the-shelf IP).

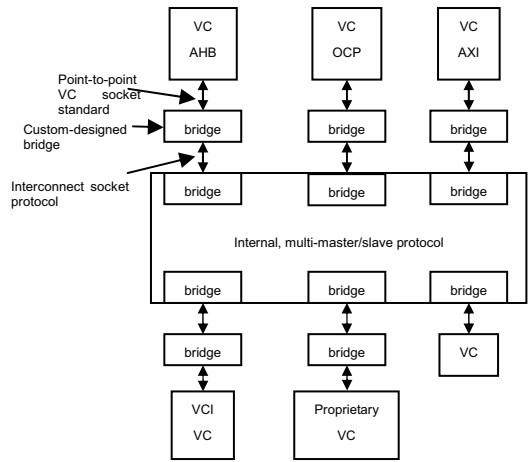

**Fig 2: Usual system with mixed VC protocols**

Bridges introduce area and latency penalties, but worse, they also do not support the full set of VC transactions because they are limited by the interconnect protocol and physical design. Compared to this approach, the layered architecture provides a strong advantage to NoCs with respect to compatibility handling:

- The switching fabric managing transport being transaction-unaware, transactions can be customized to the actual set of VCs that plug into the NoC, without altering the transport and physical layers.

- Adding a socket-specific feature to a NIU is easy because the NoC protocol layering drives the process with a simple set of questions:

1. Does the feature require some specific transaction state to be stored in the NIU? If yes, add the state to the standard NIU state lookup tables (which track for example that a Load request is waiting for a response).

2. Does it require information to be exchanged between NIUs? If yes, add it to the packet format.

Since neither adding bits to the packets nor state in the NIUs impacts transaction or physical layers, supporting VC-specific features in the NoC only impacts the corresponding NIU. However, the various VC sockets have many incompatibilities in their basic features that need to be carefully addressed upfront in the NoC transaction layer.

## 3. Incompatibility examples and solutions

Many VC compatibility issues must be carefully analyzed and resolved to reach a truly VC neutral NoC. Among these are: ordering model, exclusive accesses and atomicity, bursts, endianness. We will discuss ordering model and exclusive accesses below.

AHB, PVCI, BVCI VCs are fully-ordered between requests and responses. OCP compatible VCs are fully ordered within a thread, but may be multi-threaded with no ordering constraint between threads. AXI and AVCI provide transaction IDs with requests and response, also allowing out-of-order responses. Some of these protocols (such as OCP) support WRITEs without responses, and others (such as AXI) have independent READ and WRITE channels, further obscuring ordering constraints.

The various flavors of ordering models and pipelining schemes between these protocols create a challenge to define a transaction layer compatible with all configurations, while keeping a low NIU gate count.

Arteris NoC protocol uses a packet destination field (called SlvAddr), a packet source (called MstAddr), and a Tag. The ordering model adapts to the fully-ordered AHB, the multi-threaded OCP and the ID-based AXI ordering models using a careful assignment policy of these fields from the OCP or AXI ones such as ThreadID and TID. Further, this policy is flexible and allows NIUs to support one or many simultaneously outstanding transactions and/or targets, scaling their gate count to their expected performance within the system.

As outlined above, the NoC switch fabric itself is unaware of actual NIU field assignment policies – it sees uniform packets with SlvAddr, MstAddr, Tag information, allowing the seamless plugging of OCP, AXI, AHB, etc. IP blocks to the same NoC.

OCP and AXI have introduced new transactions called « lazy synchronization » and « exclusive access » respectively. They implement non-blocking synchronization between several masters, contrary to the older READEX and LOCK transactions. To implement all these in the NoC, we found out that READEX/LOCK transactions impact transport level (switches take specific decisions when they see LOCK-related packets) but that handling AXI and OCP exclusive access only requires adding a single user-defined bit in the packets, and state information in the NIU. This optional packet bit becomes simply part of a family of similar "NoC services" that can be activated in a particular NoC configuration.

## 4. Summary

Arteris NoC transaction layer is compatible with AHB, VCI flavors, AXI, OCP transaction levels and thus allows seamless mixing of IP blocks using these sockets. The layered structure of NoCs allows IP compatibility to be independent from transport and physical layer specification, allowing easy plug-in of new socket protocols to the NoC through appropriate NIUs, while at the same time the transport and physical layers can be adapted to specific QoS requirements or physical environments.